\documentclass{article}

\input{tcilatex}
\begin{document}

\title{A Note on Entanglement, Nonlocality, and Superluminal Signaling%
\thanks{%
Dedicated to the hippies who saved physics, despite not sharing their
interest in paranormal phenomena and extra sensorial perception: \textit{How
the Hippies Saved Physics: Science, Counterculture, and the Quantum Revival}%
, by David Kaiser (W. W, Norton \& Company, 2012)}}
\author{Luiz Carlos Ryff \\
\textit{Instituto de F\'{\i}sica, Universidade Federal do Rio de Janeiro, }\\
\textit{Caixa Postal 68528, 21041-972 RJ, Brazil}\\
E-mail{\small : ryff@if.ufrj.br}}
\maketitle

\begin{abstract}
An alternative approach to \textquotedblleft entanglement, nonlocality, and
superluminal signaling\textquotedblright\ is advocated. It is inspired on
Faraday's way of viewing physics.
\end{abstract}

In a recent and interesting article Ghirardi discusses different proposals
of superluminal signaling, proving why they do not work \textrm{[1]}. In
fact, Bell's theorem and Einstein-Podolsky-Rosen (EPR)\ correlations\ 
\textrm{[2] }strongly suggest that (using Bell's own words)
\textquotedblleft behind the scene something is going faster than
light\textquotedblright\ \textrm{[3]}, and Bohm even entertained the idea of
superluminal signaling \textrm{[4]}. Despite the \textquotedblleft pacific
coexistence\textquotedblright\ between quantum nonlocality and special
relativity \textrm{[5]}, the former seems to go against the spirit of the
latter. So, it does not come as a surprise that now and then a new (although
unfounded) proposal for a superluminal telegraph is presented. However, the
issue is not closed. The point is in the word \textquotedblleft
if\textquotedblright . As shown in \textrm{[6]}, if EPR correlations result
from finite-speed causal influences then superluminal signaling is possible,
at least in principle. Although this seems to clash with special relativity,
things are not that simple, as shown in \textrm{[7]}.

Actually, to be honest, it is important to stress that not all physicists
agree on this matter. For some, the so called collapse of the state vector,
supposed to be responsible for the nonlocal features of EPR correlations, is
a purely subjective process, occurring only inside the mind of the observer 
\textrm{[8]}. However, in my opinion, this position, as others that purport
to dismiss quantum nonlocality, is untenable. Rejecting unlikely
coincidences, supernatural causes, fatalism, and actions from the future
into the present,\textrm{\ }there are two possible explanations for the
observed correlations between distant events: a previously shared property
or some kind of interaction. As shown by Bell \textrm{[2]}, the first cannot
account for EPR correlations. We must then maintain the second (obviously,
the simply abandon of realism cannot explain the correlations either \textrm{%
[9])}. Although it may \ sound trivial or much too obvious, it is worth
remembering that in an experiment to test Bell's inequalities the data are
automatically registered. The role of the observer is to see what has been
recorded and to verify that, according to Bell's theorem, no property
previously shared by the particles can explain the correlations \textrm{[10]}%
.\textrm{\ }

Naturally, once we accept that the forcing of the particle of an entangled
pair into a well-defined physical state can alter the state of the other,
distant one, some questions may be raised. In the case of events separated
by a space-like interval it is not possible to know which one took place
first, triggering the superluminal causal influence. This strongly suggests
the existence of a preferred frame, reminiscent of Newton's absolute space,
in which the real time sequence of events would be known. As has been argued 
\textrm{[11]}, this does not imply that the well-verified results derived
from the assumption of Lorentz covariance have to be abandoned. We may also
conjecture about the speed of this faster-than-light (FTL) interaction ($v_{%
\text{{\tiny FTL}}}$) \textrm{[12]}. If the connection between the entangled
particles takes place in our ordinary 3-dimensional space, then we must have 
$v_{\text{{\tiny FTL}}}\neq \infty $, since, strictly speaking, infinite
does not correspond to a definite value. On the other hand, if \ $v_{\text{%
{\tiny FTL}}}=\infty $, even distant from each other the particles must
constitute a unique single system, and our customary notion of space has to
be revised.\textrm{\ }Of course, a connected question is related to the time
duration of the collapse-inducing process. To try to be more precise:
Exactly what kind of process induces collapse? Furthermore, \textit{when}
and \textit{where} can measurement be considered accomplished, and what
makes measurement different from other physical processes? As time has
shown, these are not straightforward questions, and no consensus has yet\
been reached regarding them.

Things may become even more blurred when null-result (NR) (or negative)
detections are considered: when a detector does not click, can a collapse of
the wavefunction still take place \textrm{[13]}? An affirmative answer
implies that no irreversible amplification is needed to induce the reduction
of the state vector. Then, we may conjecture that at a\textrm{\ }deep and
fundamental level some as yet unknown processes take place which are
responsible for the so called \textquotedblleft actualization of
potentialities.\textquotedblright\ Although NR measurement is a relatively
old subject, in my opinion it still needs to be more fully experimentally
investigated \textrm{[7, 13]}. As has been shown \textrm{[7, 14]}, if NR
detections do not have the same capability of reducing (or collapsing) the
quantum state vector as ordinary detections then superluminal signaling
becomes possible, as entertained by Bohm \textrm{[4, 7]}.

Another important question is: How is this possible information about the
outcome of a measurement performed on a photon of an entangled pair conveyed
to its distant twin? A tentative mechanism that can be experimentally tested
has been suggested (Ryff, in ref. \textrm{[6]}).\textrm{\ }Probably, we need
a more intuitive, Faraday-like, approach to quantum nonlocality in order to
disclose new possibilities \textrm{[15]}.

\end{document}